\documentclass[a4paper]{jpconf}
\usepackage{amsmath,amssymb,graphicx}
\usepackage[bookmarks=false]{hyperref}
\pdfoutput=1

\def\bm#1{\mbox{\boldmath$#1$\unboldmath}}
\DeclareMathOperator{\diag}{diag}
\newcommand{\Mkk}{M_{\rm KK}}

\newcommand{\ie}{{\it i.e.}}
\newcommand{\eg}{{\it e.g.}}

\begin{document}

\title{Indirect tests of the Randall-Sundrum model}
\author{Sandro Casagrande}
\address{Excellence Cluster Universe,
  Technische Universit\"at M\"unchen \\
  D-85748 Garching, Germany}
\ead{sandro.casagrande@ph.tum.de}
\begin{abstract}
  I present phenomenological implications of the Randall-Sundrum model
  for indirect searches, specifically a selection of flavor
  observables and Higgs-related collider searches. I review the
  interplay of constraints from CP violation in flavor physics,
  possible effects in rare decays, and model-specific protection
  mechanisms. Deviations in the Higgs couplings to fermions and, at
  one-loop, to gluons are unexpectedly strong and lead to strong
  modifications in Higgs searches.
\end{abstract}

\section{Introduction}

The gauge hierarchy problem is one of the driving theoretical reasons
for the invention and necessity of new physics at the electroweak
scale. Its strength can be quantified as a technical naturalness
problem induced by the instability of the electroweak scale under
radiative corrections, which gains severity given the hierarchy of 16
orders of magnitude between the electroweak and the Planck scale
$M_{\rm PL}$.  Models with compact extra dimensions explain this
hierarchy in terms of geometry, and at the same time also the
hierarchical structures observed in the fermionic masses and mixing
angles via so-called geometrical sequestering
\cite{ArkaniHamed:1999dc}.

This can be achieved naturally within the framework of a warped extra
dimension, first proposed by Randall and Sundrum (RS)
\cite{Randall:1999ee}.  There one studies the Standard Model (SM) on a
background consisting of Minkowski space, embedded in a slice of
five-dimensional anti de-Sitter geometry ($AdS_5$) with curvature
$k$. The fifth dimension is an $S^1/Z_2$ orbifold of size $r$, and has
two branes located at orbifold fixed points, the UV and the IR
brane. The geometry is given by
\begin{equation}
  ds^2 = e^{-2\sigma(\phi)}\,\eta_{\mu\nu}\,dx^\mu dx^\nu
    - r^2 d\phi^2 \,,\quad \sigma(\phi)=kr|\phi| \,,
\end{equation}
where $e^{\sigma(\phi)}$ is called the warp factor.
% A negative bulk cosmological constant is balanced by tensions on the
% two branes, which makes it solve the 5d Einstein-equations.
% Also simple explicit setups have been constructed to generate a
% radion potential \cite{Goldberger:1999uk} stabilizing the size of
% the extra dimension.
For $L \equiv kr\pi \equiv - \ln(\epsilon) \approx \ln(10^{16})$ the
exponential contraction of length scales along the extra dimension
mediates between a physical scale $\Mkk \equiv k' \equiv k\epsilon
\sim \text{TeV}$ and a fundamental scale $k \lesssim M_{\rm Pl}$. We
use the common setup of allowing the gauge
% \cite{Davoudiasl:1999tf, Pomarol:1999ad, Chang:1999nh}
and matter fields
% \cite{Grossman:1999ra, Gherghetta:2000qt}
to spread in the $AdS_5$ bulk. This SM-like 5d gauge theory is the
minimal setup from the dual conformal field theory
% \cite{Maldacena:1997re}
point of view, as it implies that only the Higgs is a total TeV-scale
composite of the strongly interacting sector, while SM gauge fields
are fundamental and SM fermions are mostly fundamental with small
mixing to the conformal sector \cite{Agashe:2003zs}. Pragmatically
thought, we only protect the scalar mass, leaving masses of
higher-spin fundamentals to be protected by chiral or gauge
symmetries.\footnote{For a more exhaustive review on the development
of the model see \cite{Casagrande:2008hr} and the references therein.}

Fermions in the bulk admit a natural explanation of the flavor
structure of the SM \cite{Grossman:1999ra, Gherghetta:2000qt,
Huber:2003tu}. We achieve this by exponential localization of the
fermion zero modes, which are the lowest mass harmonics of the 5d
fermions. They receive their physical mass only from effective Yukawa
couplings, resulting from their wave-function overlap with a Higgs
boson. The latter is confined very close to the IR brane, where we may
use the limit of complete localization for simplicity. The fermionic
IR brane value $F(c)$ is exponentially suppressed by the volume factor
$L$, if the normalized bulk mass parameters $c_{Q_i}=+M_{Q_i}/k$, and
$c_{q_i}=-M_{q_i}/k$ are smaller than a critical value
$-1/2$. Otherwise $F(c)$ is of order one. Here $M_{Q_i}$ and $M_{q_i}$
denote the masses of the five-dimensional $SU(2)_L$ doublet and
singlet fermions. The exponential hierarchies are used to generate a
hierarchical Yukawa matrix out of anarchical $\mathcal{O}(1)$
fundamental Yukawa coupling
\begin{equation}
  \bm{Y}_q^{\rm eff} = \diag\big( F(c_{Q_i}) \big) \bm{Y}_{\!q} \,
  \diag\big( F(c_{q_i}) \big) \,,\quad F(c) \approx \exp\!\big(\!
  \min(c+1/2, 0) L \big) \sqrt{1+2c} \,.
\end{equation}

\begin{figure}[h] \begin{center} \vspace{-2mm}
    \includegraphics[width=15cm]{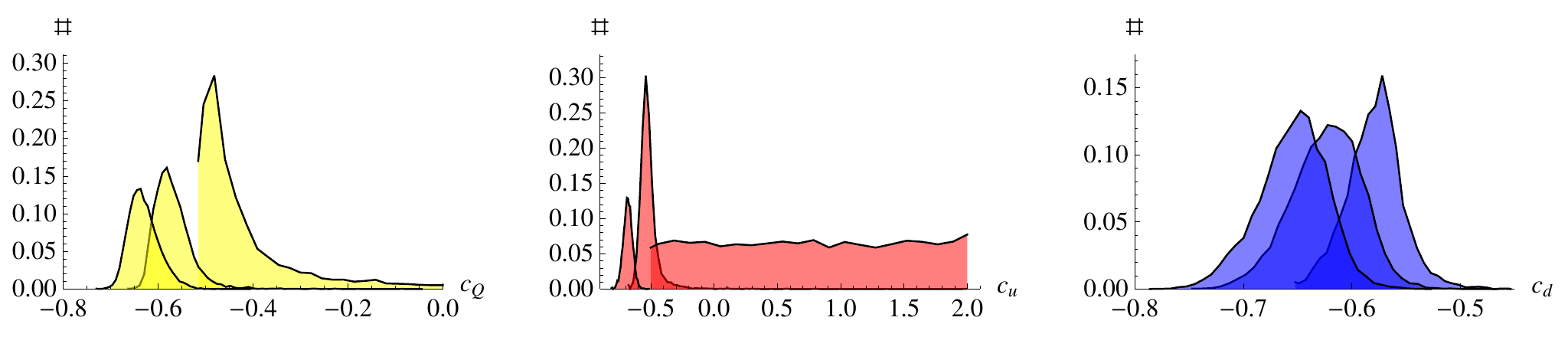} \vspace{-2mm}
    \caption{\label{fig:c_distribution} Distribution of bulk masses
      for $\Mkk \in [1,10]$ TeV, $\epsilon = 10^{-16}$, and flat
      $c_{t_R}$ prior.} \vspace{-4mm}
\end{center} \end{figure}

The model features a plethora of higher harmonics above the TeV scale,
the KK excitations of gauge bosons starting at $m_{g^{(1)}} \!\approx
2.45\, \Mkk$, graviton $m_{G^{(1)}} \!\approx 3.83\, \Mkk$, etc. each
followed by an infinite tower of states, and additional mixings with
the SM fields. Yet it is determined by a rather small set of new
parameters \cite{Agashe:2004cp}, \eg\ in the quark sector we have $18$
additional moduli and $9$ phases.  Moreover, explaining the SM fermion
spectrum constrains all ratios of the values $F(c)$. E.g. in the quark
sector \cite{Casagrande:2008hr, Huber:2003tu, Blanke:2008zb}, up to
$\mathcal{O}(1)$ factors
\begin{equation}
  m_{q_i} \sim \frac{v}{\sqrt{2}} F(c_{Q_i}) F(c_{q_i}) \,,\quad
  \lambda \sim \frac{F(c_{Q_1})}{F(c_{Q_2})} \,,\quad A \lambda^2
  \sim \frac{F(c_{Q_2})}{F(c_{Q_3})} \,,
\end{equation}
leading to the numerical distribution shown in figure
\ref{fig:c_distribution}, where we used exact formulas
\cite{Casagrande:2008hr}.
Thus the model has well testable predictions, which could be observed
in future measurements, yet we have to assure conformity with existing
measurements. Most prominently, the compliance with electroweak
precision data tightly restricts the minimal model to $\Mkk > 4.0$ TeV
at 99\% CL \cite{Carena:2003fx, Casagrande:2008hr}. There exist
several ways to relax this constraint, like a possible high Higgs mass
of about $1$~TeV, which would be natural in this model and cancels the
model-dependent contribution to the $T$ parameter, leading to a lower
bound $\Mkk > 2.6$ TeV. Also restricting the model to warp only up to
an intermediate and safe UV scale can help to lower the bound (see
\cite{Bauer:2008xb} for restrictions of this approach). Alternatively
one can make use of the custodial $O(4)$ symmetry of the Higgs sector,
by extending the bulk hypercharge group to $SU(2)_R\times U(1)_X$,
breaking it to $U(1)_Y$ on the UV brane by specific orbifold boundary
conditions \cite{Agashe:2003zs}. The Higgs mechanism on the IR brane
may safely break $SU(2)_L \times SU(2)_R$ to the diagonal $SU(2)_V$
subgroup. This suppresses tree level contributions to $T$ by three
orders of magnitude and implies $\Mkk > 2.4$ TeV, which can even be
lowered further by calculable fermionic loop contributions to
$T$.\footnote{However, quark bi-doublets necessary to protect
  $Zb_L\bar{b}_L$ typically tend to give a negative contribution to
  $T$, which would even raise the bound \cite{Carena:2007ua}.} A
discrete left-right parity symmetry and specifically chosen fermion
representations \cite{Agashe:2006at} furthermore relax constraints
from the precisely measured $Zb\bar{b}$ vertex allowing for a wide
range of right handed bottom and left handed top/bottom bulk masses,
even for a low new physics scale.

It is well known that strong constraints on the parameter space of new
physics models can also arise from flavor observables. Depending on
the flavor structure and the first quantum order of flavor changing
neutral currents, this translates into a bound on the new physics
scale, and might result in a tension to the resolution of the
hierarchy problem. This is the so called flavor puzzle, which is
partly addressed in the RS model in the way fermion hierarchies are
generated. The mechanism is referred to as the RS-GIM and we show its
numerical implications below.

Given the relatively high bounds on the scale $\Mkk$ and consequently
on new particles, we consider it worthwhile to present possibilities
which might indirectly allow for detection or exclusion of a warped
extra dimension. We start at the high luminosity frontier with effects
on flavor observables, emphasizing the most prominent bound coming
from indirect CP violation (CPV) in $K$--$\bar{K}$ transitions. We
show that also direct CPV provides an effective constraint on some
rare decays and discuss selected correlations, based on the exhaustive
survey \cite{Bauer:2008xb}. At the high energy frontier we present a
recent full one-loop calculation of the effect in Higgs production and
decays \cite{Casagrande:2010si}, which shows in strong shifts in the
feasibility of Higgs searching strategies.

Since the RS-GIM mechanism implies flavor dependent couplings ordered
with the fermion masses, it is most reasonable to expect changes in
top couplings. This could affect for instance the forward-backward
asymmetry in $t\bar{t}$ production, for which the measurement by CDF
has very recently been affirmed \cite{Aaltonen:2011kc}, and a
significant deviation from the NLO QCD prediction in the high
invariant mass region has been found. However one must carefully
account for the explanation of the flavor structure. The latter
renders the leading-order axial-vector couplings to the KK-gluon
dominated contributions far too small to explain the discrepancy, and
also the interference of vector couplings with the one-loop SM
diagrams is outweighed by the leading-order contribution to the
symmetric cross section $\sigma_{t\bar{t}}$ \cite{Bauer:2010iq}.
Another coupling of the top quark, namely to a Higgs boson and a charm
quark, receives enhanced importance \cite{Azatov:2009na} for which we
show numerics to round off this presentation.

\section{Flavor related effects and constraints}

Indirect CPV in $K \rightarrow \pi \pi$ decays has an
outstanding constraining power in many extensions of the SM. It is
induced by $K - \bar{K}$ transition and the most recent theoretical
determination of the imaginary value of the corresponding transition
amplitude is given by $\epsilon_K = (1.90 \pm 0.26) \cdot 10^{-3}$
\cite{Brod:2010mj}.  In the effective-theory parametrization
\begin{equation}
  \mathcal{H}_{\rm eff}^{\Delta S=2} = \sum_{i=1}^5 C_i\,Q_i^{sd} +
  \sum_{i=1}^3 \tilde C_i\,\tilde Q_i^{sd} \,,
\end{equation}
helicity structures beyond the $V\!-\!A$ couplings of charged SM
currents can contribute to the amplitude. Most importantly $Q_4 =
(\bar d_R s_L)\,(\bar d_L s_R)$, can be generated by the exchange of
new colored particles. Its coefficient enters the approximate formula
for the $K$--$\bar K$ mixing amplitude
\begin{equation}
  \left \langle K^0 \left | {\cal H}_{\rm eff, RS}^{\Delta S = 2}
    \right | \bar K^0 \right \rangle \, \propto \, C_1^{\rm RS} +
  \widetilde C_1^{\rm RS} + 114.8 \, \bigg (1 + 0.14 \, \ln \left (
    \frac{\mu_{\rm KK}}{3 \, {\rm TeV}} \right ) \! \!  \bigg ) \left(
    C_4^{\rm RS} + \frac{C_5^{\rm RS}}{3.1} \right) \,.
\end{equation}
We observe that $C_4^{\rm RS}$ has a large coefficient, which
originates from the chiral enhancement of the hadronic matrix
elements, and % a factor of about 8 due to
the RG evolution % from $\mu_{\rm KK} = 3 \, {\rm TeV}$
down to $2 \, {\rm GeV}$. Thus the strongest contribution comes from
KK-gluon exchange \cite{Blanke:2008zb, Csaki:2008zd, Bauer:2009cf}.
The RS-GIM is at work, as each fermion attached to a flavor violating
vertex contributes approximately a factor of its IR brane value $F(c)$
leading to $C_4^{\rm RS} \sim \frac{8 \pi \alpha_s}{ \Mkk^2} \, L \,
F(c_{Q_1}) F(c_{Q_2}) F(c_{d_1}) F(c_{d_2})$.

\begin{figure}[t] \vspace{-5mm} \begin{center}
    \includegraphics[width=7cm]{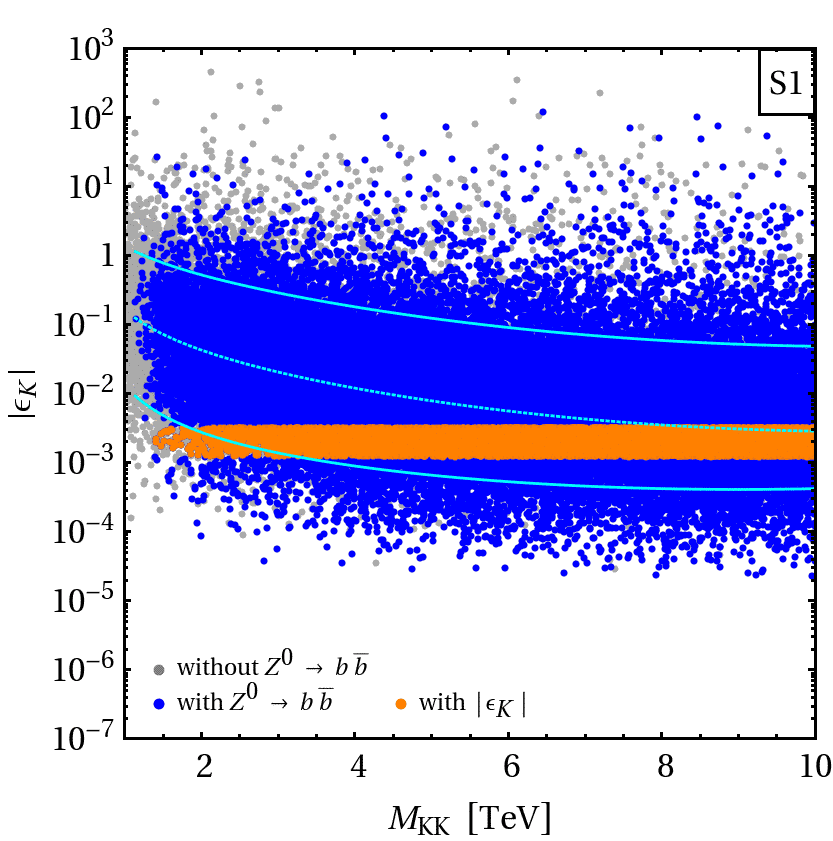} \hspace{10mm}
    \begin{minipage}[b]{6.5cm}
      \caption{\label{fig:epsK} Prediction for $|\epsilon_K|$ in the
        setup with minimal field content. All points reproduce the observed
        fermion spectrum, and CKM matrix, are not excessively fine-tuned,
        %(less than $10^3$)
        and have perturbative Yukawa couplings up to the $2^{nd}$
        KK-level. Points colored blue are consistent with $Z^0 \to b
        \bar b$ at the 99\% CL, and orange points fall inside the
        experimental 95\% CL of $|\epsilon_K|$ combined with the
        theory error. The cyan lines illustrate the decoupling
        behavior obtained from a fit to the $5\%$, $50\%$, and $95\%$
        quantile.\vspace{3mm}}
    \end{minipage}
\end{center} \vspace{-6mm} \end{figure}

In figure \ref{fig:epsK} we observe that the median value of
$|\epsilon_K|$ is consistent with the measurement only for $\Mkk
\gtrsim 8$ TeV. However the $5\%$ quantile crosses the experimentally
allowed range already at about $2$ TeV. The slow decoupling therefore
introduces a moderate amount of arbitrariness in the definition of a
bound on the new physics scale inferred from this observable, and it
is important to be aware of the different quality compared to the
aforementioned bounds from electroweak precision observables.

One possibility to protect $|\epsilon_K|$ from excessive corrections
is to arrange for common bulk mass parameters in
the sector of the right-handed down-type quarks
\cite{Santiago:2008vq}. In \cite{Bauer:2009cf} we showed that the
typical suppression of $C_4^{\rm RS}$ amounts to a factor of $8 \cdot
10^{-3}$ and discuss a few other resolutions.

\begin{figure}[b] \vspace{-4mm} \begin{center}
    \includegraphics[width=7cm]{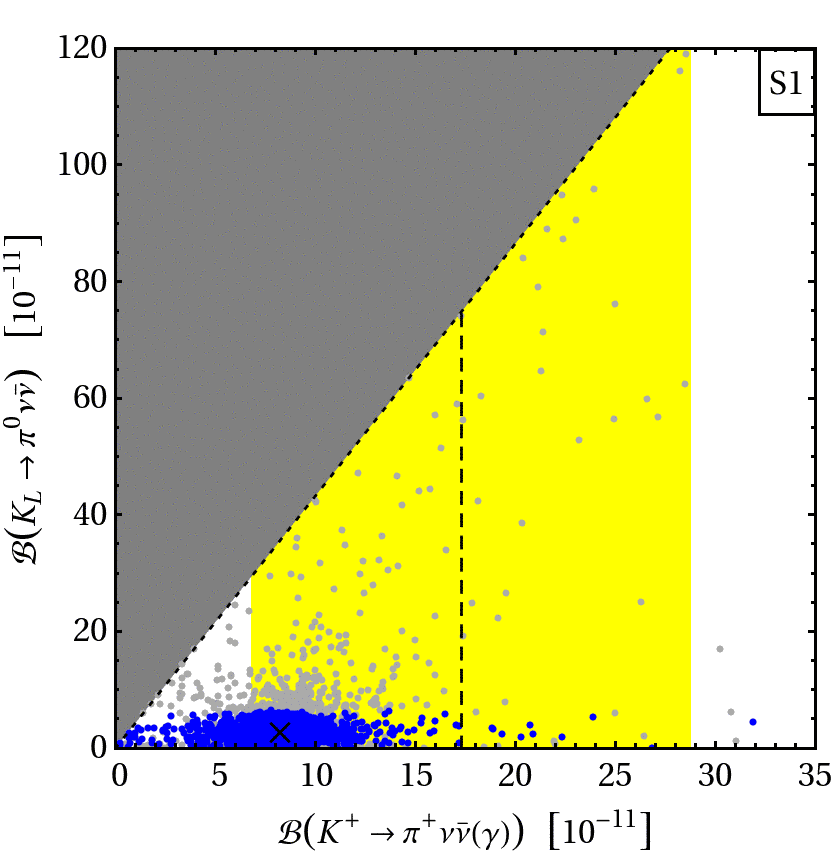} \hspace{10mm}
    \begin{minipage}[b]{6.5cm} 
      \caption{\label{fig:Kpinunu_with_epseps} Impact of the
        $\epsilon_K^\prime/\epsilon_K$ constraint on $\mathcal{B}(K^+
        \to \pi^+ \nu \bar \nu (\gamma))$ and $\mathcal{B}(K_L \to
        \pi^0 \nu \bar \nu)$. All points reproduce the quark masses
        and mixings, and the measured value of $|\epsilon_K|$,
        observables in $Z^0 \to b \bar b$, and $B_d$--$\bar B_d$
        mixing. The blue points are consistent with the measured value
        of $\epsilon_K'/\epsilon_K$ after varying the hadronic
        parameters. For details see \cite{Bauer:2009cf}.
        The black cross indicates the SM point, the light gray shaded
        area shows the region excluded by isospin conservation of QCD,
        and the yellow band displays the experimental 68\% CL
        range. \vspace{0mm}}
    \end{minipage}
\end{center} \vspace{-5mm} \end{figure}

The hadronic uncertainties involved in direct CPV in $K \to \pi\pi$
are larger compared to those in $\epsilon_K$. Yet it turns out
to be useful to look at $\epsilon_K'/\epsilon_K$, which describes the
ratio of direct over indirect CPV in this decay and is defined in
terms of ratios of $\Delta I = 3/2$ and $\Delta I = 1/2$ transitions
\begin{equation}
  \frac{\epsilon_K'}{\epsilon_K} = \frac{1}{\sqrt2}\left(\frac{A(K_L
      \to (\pi\pi)_{I_2})}{A(K_L \to (\pi\pi)_{I_0})} - \frac{A(K_S \to
      (\pi\pi)_{I_2})}{A(K_S \to (\pi\pi)_{I_0})} \right) \,.
\end{equation}
Taking into account the SM Wilson coefficients and the
renormalization-group evolution from $\mu_W$ to $\mu_c$
\cite{Buras:1993dy, Bosch:1999wr} we observe that contributions to the
QCD penguin % $Q_6$
and the electroweak penguin % $Q_8$
\begin{equation}
  Q_6 = 4\,(\bar s_L^\alpha\gamma^\mu b_L^\beta) \sum{}_{\!q}\, (\bar
  q_R^\beta\gamma_\mu q_R^\alpha) \,,\quad Q_8 = 6\,(\bar
  s_L^\alpha\gamma^\mu b_L^\beta) \sum{}_{\!q}\, Q_q\,(\bar
  q_R^\beta\gamma_\mu q_R^\alpha) \,,
\end{equation}
play the dominant role \cite{Bauer:2009cf}. Indeed, both QCD and
electroweak penguin contributions generated at a high scale are
strongly enhanced by the RG evolution from $\mu_W$ down to
$\mu_c$. However, QCD corrections mainly result from the mixing of
$Q_6$ with current-current operators $Q_{1,2}$, and are therefore
mostly unaffected by the given high-scale physics. Concerning the
electroweak side, in the RS model only contributions to the
color-singlet operator $Q_7$ arise at the scale $\mu_{\rm KK}$, but
they are directly fed via operator mixing % with a factor of $1.72$
into $C_8$ at the scale $\mu_W$. Therefore $\epsilon_K'/\epsilon_K$ in
RS is mainly influenced by the electroweak penguin, similar to the
decays $K \to \pi \nu\bar\nu$ where they are the only
contributions. For those very decays of charged and neutral Kaons the
SM amplitude is tiny, both because of the $V_{ts}^* V_{td}$
suppression induced by the CKM structure, and the GIM mechanism
\cite{Buchalla:1998ba}.  The necessary low energy matrix elements for
the decays are related to precisely measured semileptonic tree-level
decays $K_{l3}$. Therefore they offer a very clean test of physics
beyond the SM from the theoretical perspective \cite{Brod:2010hi}, but
are likewise challenging to measure.  Fortunately two experiments aim
to perform measurements with an anticipated accuracy of $10\%$, namely
NA62 (CERN) for the charged mode and Koto (J-PARC) for the neutral
mode.

In figure \ref{fig:Kpinunu_with_epseps} we observe that the RS and SM
amplitudes are comparable in size. Indeed, the factors
$F(c_{Q_1})F(c_{Q_2})$ associated with the flavor-violating RS vertex
account for approximately the same suppression of $\lambda^5$ as in
the CKM structure of the SM vertex. Moreover, the relevant RS
corrections to the left-handed $s\rightarrow dZ^0$ amplitude have a
free phase, in contrast to the small phase in the SM. Since the CP
violating neutral mode is proportional to the imaginary part of the
contributions, we see that it could in principle obtain large relative
corrections, which are however excluded by the following argument. The
typical ratio of RS over SM contributions in $\epsilon_K'/\epsilon_K$
is even a factor of $\sim 20$ larger than in $K \to \pi \nu \bar\nu$
case. The reason is well known: The QCD and electroweak penguins
cancel in the SM to a large extent, so the value of
$\epsilon_K'/\epsilon_K$ is accidentally small. This entails an
effective constraint. The RS amplitude adds linearly and with opposite
sign to the SM contribution in $\epsilon_K'/\epsilon_K$, and
quadratically with the same sign in $K_L \rightarrow \pi^0 \nu
\bar\nu$. As the central value of theory prediction for
$\epsilon_K'/\epsilon_K$ is below the experimental one, large
enhancements of $K_L \rightarrow \pi^0 \nu \bar\nu$ are disfavored
already with a conservative treatment of hadronic errors.  A lattice
determination of the hadronic matrix elements of $K \rightarrow
\pi\pi$, taking into account the full final state, is in progress
\cite{Liu:2010fb}. We emphasize that this would be most welcome in
view of the constraining potential.

\begin{figure}[h] \vspace{-1mm} \begin{center}
    \includegraphics[width=7cm]{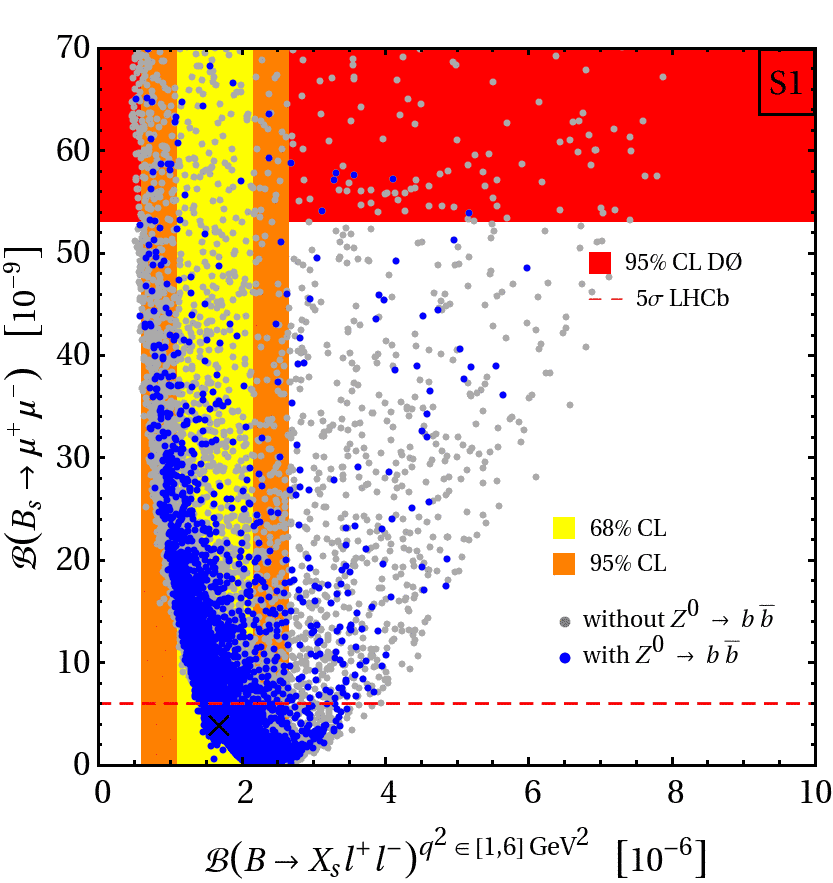} \hspace{10mm}
    \begin{minipage}[b]{6.5cm}
      \caption{\label{fig:BXsll} Correlation of the prediction for
        $\mathcal{B}(B \to X_s l^+ l^-)$ versus $\mathcal{B} (B_s \to
        \mu^+ \mu^-)$ in the setup with minimal field content. The
        blue points are consistent with the measured $Z b \bar
        b$ couplings at the 99\% CL. The black cross indicates the SM
        expectation.  For comparison the regions of 68\% (yellow) and
        95\% (orange) CL are also displayed. Furthermore the 95\% CL
        exclusion of $\mathcal{B}(B_s \to \mu^+ \mu^-)$ and the
        minimum branching fraction allowing for a discovery with
        $5\sigma$ at LHCb are indicated by the red band and dashed
        line.\vspace{5mm}}
    \end{minipage}
\end{center} \vspace{-3mm} \end{figure}

To briefly show one of the many correlations in the $B$-meson sector,
we select here the decays $\mathcal{B}(B \to X_s l^+ l^-)^{q^2 \in [1,
6] \, {\rm GeV}^2}$ and $\mathcal{B}(B_s \to \mu^+ \mu^-)$ given in
figure \ref{fig:BXsll}.
After imposing the $Z b \bar b$ constraint, enhancements in the purely
leptonic mode imply values of $\mathcal{B}(B \to X_s l^+ l^-)^{q^2 \in
[1, 6] \, {\rm GeV}^2}$ safely within the experimentally allowed range
in most of the parameter space. This is expected in models where third
generation couplings are most strongly modified \cite{Haisch:2007ia}.
The anti-correlation can be understood by the fact that both modes
receive the dominant contribution from axial vector couplings, which
are aligned in flavor space, whereas the corresponding SM amplitudes
have opposite sign in the two decays. Leptonic $B$-meson decays belong
to the channels that can be studied by three LHC experiments, ATLAS,
CMS, and LHCb. They will probe the branching fraction of $B_s \to
\mu^+ \mu^-$ down to its SM value and might reveal a signal of new
physics well ahead of the direct searches.

We conclude the discussion of flavor observables with a final remark
on the changes that occur when the minimal bulk gauge symmetry is
extended to a custodial symmetry with left-right parity as mentioned
in the introduction. The changes in the couplings have been worked out
\cite{Casagrande:2010si} and the leading effects encoded in
flavor-universal rescaling factors. Custodial RS effects in $Z d^i_L
\bar d^j_L$ (and $Z u^i_R \bar u^j_R$, $W u^i_L \bar u^j_L$) couplings
are reduced by $1/L$, but at the same time increased by a factor of
approximately $9$ in $Z d_R^i \bar d_R^j$ relative to the minimal RS
model. Effects in $K\rightarrow\pi\nu\bar\nu$ decays are therefore
typically reduced if $t_R$ has a natural bulk mass, and large
enhancements of $B_{s,d} \rightarrow \mu^+ \mu^-$ like shown in figure
\ref{fig:BXsll} are impossible \cite{Blanke:2008yr}. The change of
the chiral nature of the $Zsd$ vertex remarkably inverts a correlation
found between the branching ratios of $K^+ \to \pi^+ \nu\bar\nu$ and
$K_L \to \mu^+ \mu^-$, but an identification of the chiral structure
in this way would require major theoretical progress in the prediction
of this radiative decay.

Similarly one can work out the relative coupling shifts for $\Delta F
= 2$ processes, which shows that $Z'$ exchange is modified by a factor
of approximately $-12$ in $C_5$. Although KK-gluon exchange is
still dominant in $K$--$\bar K$ transitions, the $Z'$ can be
competitive in $B_{q}$--$\bar B_{q}$ mixing.

\section{Consequences for Higgs production and decays}

% In for one generation even in mass eigenstates it is feasible to do
% the sum analytically. The geometric like series in fact give polygamma
% functions involving bulk mass parameters.
%
% \begin{figure}[htb] \begin{center}
%     \includegraphics[width=6cm]{nu_one_generation.pdf} \hspace{5mm}
%     \begin{minipage}[b]{5cm}
%       \caption{\label{fig:nu_one_generation} Example of the bulk mass
%         dependence of the sum over all KK mass-eigenstates for a
%         single fermion generation.}
%     \end{minipage}
%   \end{center} \end{figure}

\begin{figure}[b] \vspace{-1mm} \begin{center}
    \includegraphics[width=7cm]{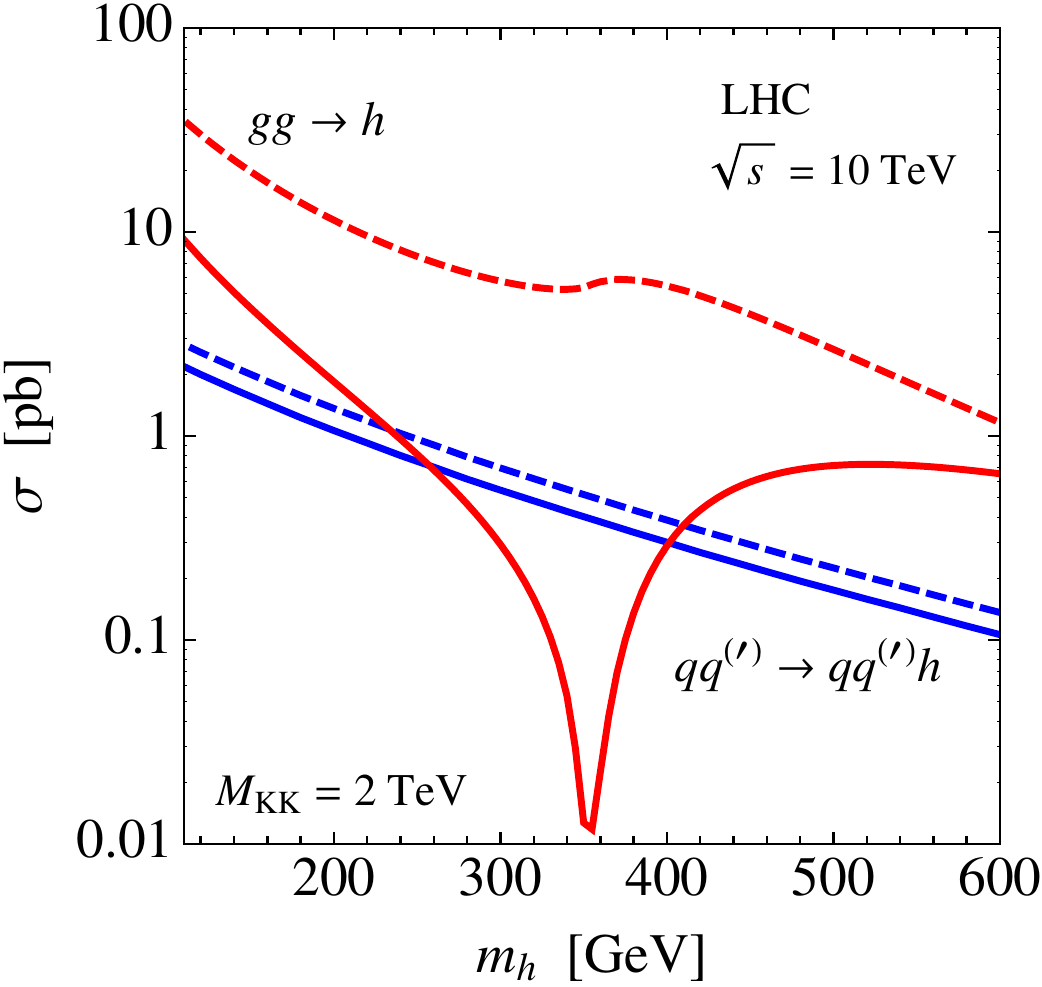} \hspace{10mm}
    \begin{minipage}[b]{6.5cm}
      \caption{\label{fig:prodplots} Cross sections for the main
        Higgs-boson production channels at the LHC for a
        center-of-mass energy of $\sqrt{s} = 10\,\text{TeV}$, and
        $\Mkk=2$ TeV. The dominant channels are gluon-gluon (red) and
        weak gauge-boson fusion (blue). The dashed lines illustrate
        the SM predictions, while the solid lines indicate the results
        obtained in the custodial RS model.\vspace{12mm}}
    \end{minipage}
\end{center} \vspace{-8mm} \end{figure}

An estimation of the impact of the RS model on Higgs searches at
hadron colliders involves first of all the production, where gluon fusion is
the main channel. In the SM, the process receives the main contribution
from a top-quark triangle loop. After factoring out numerical
constants and $\alpha_s\sqrt{G_F}$ its contribution is encoded in the
form factor $A^h_q(4 m_t^2 / m_h^2)$. A small correction arises from
the bottom-quark, but since $A^h_q$ vanishes proportional to its
argument the lighter quarks are irrelevant. In the RS model the
amplitude for the top-triangle is always reduced but still
positive. The origin lies in the modified $ht\bar{t}$ coupling, which
is strongly suppressed ($\sim -25\%$ for $\Mkk = 2$ TeV) because of
the compositeness to the top, \ie\ its IR localization.

Analog to the top triangle all KK modes contribute in triangle
diagrams. Since $A^h_q \to 1$ in the limit of a very large quark mass,
the total sum over all KK fermions appears at first sight to be
divergent, as the mass splitting of the higher modes is asymptotically
equidistant. However one can show that a delicate cancellation occurs
at each KK level between the two chiralities of the vector-like
fermion associated to each SM fermion \cite{Casagrande:2010si}. As a
result the sum converges similar to a geometric series instead of
diverging like a harmonic series.

Our numerical analysis shown in figure \ref{fig:prodplots} is based on
a rescaling of the SM calculation \cite{Ahrens:2008nc}.
% which comprises the NNLO fixed order result, combined with threshold
% logarithms from soft gluon emission.
% % and terms of $(N_c \pi \alpha_s)^n$
We find that the relevant amplitudes show only a small spread when
scanning over the experimentally allowed RS parameter space. Therefore
we display only the median values. The Higgs mass
dependence can be easily understood.
% after factoring out $\alpha_s$, which is deceasing
% all the way along increasing $m_h$
Up to the $t\bar t$ threshold the form factor increases, but it
decreases above. The KK-amplitude is negative and approaches a
constant in the heavy Higgs-mass limit. Where it becomes dominant, the
full amplitude flips sign, what shows up as a complete destructive
interference in the cross section near the $t\bar t$
threshold. Numerically the suppression at the LHC ranges between
$-45\%$ to $-100\%$ depending on the Higgs mass. At $\Mkk = 3$ TeV the
suppression is at most $-90\%$ and at $5$ TeV still up to
$-40\%$. Thus we obtain a strong sensitivity of the production channel
up to high new physics scales.\footnote{Note that the numbers have
been obtained in the model with custodial gauge symmetry, where the
multiplicity of KK-fermion states is higher than in the minimal
model.} The blue curve in the plot shows the Higgs production via weak
gauge-boson fusion which is affected universally by the modified $WWh$
vertex and thus reduced by $-20\%$. The same factor applies for
associated W boson production,
% skip?
% (Higgsstrahlung), which is the only channel that in principle would
% allow for a Higgs discovery at the Tevatron,
and associated top quark pair production is suppressed by $-40\%$.

The calculation of the subsequent possible decay modes of the Higgs
involves diagrams similar to the production channel, and furthermore
gauge boson and KK-gauge boson triangle diagrams for the
$\gamma\gamma$ and $Z\gamma$ final state. The $WWh$ and $ZZh$
couplings experience only a very small reduction. Yet, we recall the
reduced production cross section, which renders a detection in the
experimentally cleanest mode $h\rightarrow Z^*Z^* \rightarrow 4l$, in
the region $m_h > 180$ GeV, to be more difficult. For a lower Higgs
mass the decay into two photons is enhanced and can overcome the
suppression of the production cross section. The product $\sigma(gg
\to h) \, \mathcal{B}(h \to \gamma \gamma)$ is almost unchanged
($+3\%$) at the scale of $\Mkk = 2$ TeV and $m_h = 120$ GeV,
and even enhanced for a lower new physics scale. However, for an
intermediate scale of $3$ TeV the cancellation in the production cross
section becomes stronger in the low $m_h$ region, and the product
drops severely to $24\%$ of the size expected in the SM.

\section{Flavor changing top decays}

We end this presentation with a surprising observation about the
possible size of flavor changing top decays. For a Higgs coupling to
fermions one usually expects factors of the corresponding fermion
masses $m_{q_i} m_{q_j}/v^2$, stemming from the Yukawa mechanism.
Assuming this for the RS model, one would expect the branching ratio
of the flavor changing decay $t\to ch$, if kinematically allowed, to
be typically two orders of magnitude smaller than for the decay $t\to
cZ$.

Models with heavy vector-like quarks can behave differently, if those
quarks mix with the SM quarks and additionally couple to the Higgs
boson with chiralities opposite to the SM, \ie\ if the Higgs couples
to a heavy left handed singlet and a heavy right handed doublet. Under
those conditions an effective operator $( H^\dagger H )( \bar Q_i H^c
u_j )$ is induced, which contributes differently to masses and Yukawa
couplings \cite{delAguila:2000rc}.

\begin{figure}[t] \vspace{-6mm} \begin{center}
    \includegraphics[width=7.5cm]{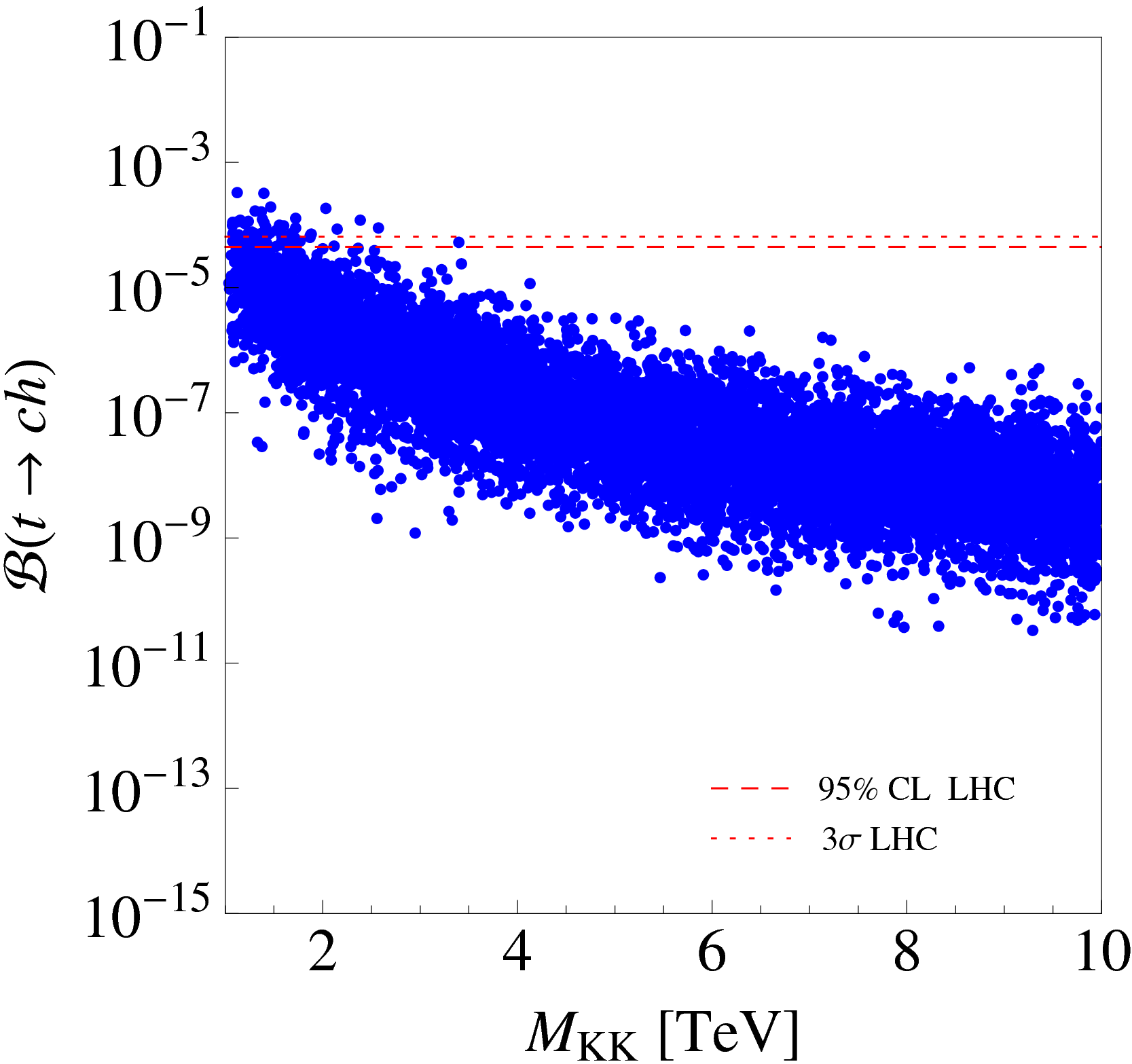} \hspace{5mm}
    \begin{minipage}[b]{6cm}
      \caption{\label{fig:tcplots} Prediction for the branching ratio
        of the rare decay $t \to c h$ in the RS model with extended
        custodial protection and $m_h = 150$ GeV.
        % $c_{{{\cal T}}_{1 i}} = c_{{{\cal T}}_{2 i}}$
        The red dotted (dashed) lines in the plot indicate the
        expected discovery (exclusion) sensitivities of LHC for
        100\,fb$^{-1}$ integrated luminosity. All points reproduce the
        correct quark masses, mixing angles, and the CKM phase. In the
        minimal RS model the constraints from $Zb\bar{b}$ typically
        eliminate points with pronounced effects.\vspace{3mm}}
    \end{minipage}
\end{center} \vspace{-6mm} \end{figure}

A correct treatment of the Higgs localization in RS shows that, even
though heavy ``wrong chirality'' fermion wave-functions are zero on
the IR brane by orbifold boundary conditions, the described type of
misalignment exists and contributes to Higgs FCNCs
\cite{Azatov:2009na}. Numerically this makes the decay $t\to ch$
typically even an order of magnitude larger than $t\to cZ$. We show
this for a reference Higgs mass $m_h = 150$ GeV in figure
\ref{fig:tcplots}, where we also indicate the experimental prospect of
detecting the process via the subsequent decay into $b\bar{b}$ + jet
\cite{AguilarSaavedra:2000aj} at later stages of the LHC running. For
a lighter Higgs mass of $m_h = 120$ GeV one multiplies the shown
branching ratio with a factor of $4.4$. Hence we conclude that
misalignment effects astoundingly might lift a detection of the $htc$
coupling into reach of the LHC.

% MSSM: Large regions of parameter space have $10^{-4}$ branching.
% General 2HDM: branching is $1.5 \cdot 10^{-3}$ by ansatz of the
%   couplings $g_{tc} \sim \sqrt{m_c m_t}/m_W$.
% For usual exotic quark models $Ztc$ should be larger than $htc$

\clearpage

\ack
It is a pleasure to thank Martin Bauer, Florian Goertz, Uli
Haisch, Matthias Neubert and Torsten Pfoh for the fruitful
collaboration and the THEP working group of the
Johannes-Gutenberg-Universit\"at Mainz for hospitality and partial
support during the writeup of this text. I am also grateful to
Joachim Brod and Martin Gorbahn for useful comments on the
manuscript. This research was supported by the DFG cluster of
excellence ``Origin and Structure of the Universe''.

\end{document}